# On the Origin of Samples:
## Attribution of Output to a Particular Algorithm

Roman V. Yampolskiy
Computer Engineering and Computer Science
University of Louisville
roman.yampolskiy@louisville.edu

**Abstract**
With unprecedented advances in genetic engineering we are starting to see progressively more original examples of synthetic life. As such organisms become more common it is desirable to be able to distinguish between natural and artificial life forms. In this paper, we present this challenge as a generalized version of Darwin's original problem, which he so brilliantly addressed in *On the Origin of Species*. After formalizing the problem of determining origin of samples we demonstrate that the problem is in fact unsolvable, in the general case, if computational resources of considered originator algorithms have not been limited and priors for such algorithms are known to be equal. Our results should be of interest to astrobiologists and scientists interested in producing a more complete theory of life, as well as to AI-Safety researchers.

## 1. Introduction

In 1859 Charles Darwin published his famous work – On the Origin of Species. In it he provided a naturalistic explanation for origins of fossilized and living biological samples collected in different regions of planet Earth. Before publication of Darwin's theory of natural selection (currently integrated into what is known as theory of evolution) the prevailing theory used to explain such samples attributed their origins to a supernatural cause commonly assumed to be god(s). Darwin's theory quickly became the dominant theory accepted by majority of scientists as the best explanation for the origins of different species. Evolutionary theory has only solidified its position over the years due to strong additional evidence from such diverse fields as genetics, anthropology, and computer science [1].

In particular, research in genetics, which was not available during Darwin's life, has provided a treasure trove of experiments used to confirm Darwin's theory. At the same time, recent unprecedented advances in genetic engineering [2], directed evolution [3] and synthetic genomics [4] have allowed scientists to create Genetically Modified Organisms (GMOs) [5], expand genetic code [6, 7], create synthetic DNA [8], and synthetic life [9] and consider creation of synthetic human genomes [10]. With the development of the latest tool for genetic manipulation (CRISPR [4]) no fundamental limits remain to engineering of novel synthetic life forms. With fields like Evolutionary Robotics [11, 12], Artificial Life [13, 14], and Evolutionary Computation [15] providing theoretical and experimental support for creation of evolvable synthetic life one can't help but think about the future directions in post-Darwinian evolutionary theory [16].

A major challenge we are likely to face in the near future is being able to tell synthetic life forms from natural ones. We are already experiencing a need to identify GMOs for proper labeling and compliance, with some early work already taking place in that domain [17-19]. With advances in space exploration, particularly with manned and unmanned spacecraft visiting moons and planets of the solar system a possibility of bacterial contamination of those space bodies by organisms from Earth becomes a real possibility. If such organisms are later rediscovered we would need to be able to determine their origin. Likewise spacecraft returning from a mission may bring unknown organisms to earth, despite our best precautions [20], again presenting us with the sample attribution problem. There is also a possibility of discovering extraterrestrial life, but we will not concentrate on this situation.

We can also setup an artificial environment in which ground truth of samples' origin and distribution is known in advance (unlike in Darwin's original problem) and attempt to select the correct explanation between modern evolutionary [21] and non-evolutionary theories [22, 23], in a side-by-side test, something we were previously not able to accomplish. Can science accurately distinguish between naturally evolved and genetically engineered life forms if the samples are known to have an equal prior distribution? This would be easy to setup in the lab once we have access to a large number of synthetic life forms. We can place 50 naturally evolved organisms (class A) and a 50 engineered organisms (class B) into a lab setting and challenge the scientific community to accurately attribute each sample with respect to class A or B. With individual samples represented by standalone artifacts, not historical records of multiple related samples. As a thought experiment we can set this up on another planet as a challenge for alien scientists/explorers to reduce any impact of knowing something about natural organisms on Earth. This produces a very clean and decisive experiment as our artificial setup removes any bias associated with results directly affecting ourselves as people on Earth and allows us to perform an experiment, the results of which can be evaluated against known truth-values. This would give us a chance to evaluate our theories of origin of biological samples.

Occam's razor [24], which states that among multiple possible hypotheses the simpler one should be selected, is typically used to argue that evolutionary theory provides a superior explanation in contrast to theories which may include an engineer as such theories have to also explain nature and origins of the said engineer resulting in a more complicated hypothesis. However, in our proposed experiment, the nature of the engineer is known and samples are chosen to have equal likelihood of being generated by evolutionary or synthetic means making application of Occam's razor erroneous.

## 2. Generalized Sample Attribution Problem

Proposed problem of telling synthetic life from naturally evolved life forms can be seen as addressing a meta-challenge of selecting the algorithm responsible for generating observed samples from a number of possible algorithms, not the original problem faced by Darwin of developing a naturalistic algorithm, which could be used to explain collected biological samples. This problem is a subset of Solomonoff Induction (SI) [25, 26] and science in general [27]. Given a set of observations, determine which of many theories best accounts for what was observed and accurately predicts future observations.

To distinguish it from Darwin's original problem let's call this problem the Generalized Sample Attribution Problem (GSAP) or Generalized Darwin's Problem. GSAP can be represented as a computer science problem, meaning in terms of algorithms and digital data. Any type of scientific samples and DNA code in particular could be represented as a bit string. Algorithms capable of generating bit strings encoding collected samples can be subdivided into two main types: evolutionary algorithms (Genetic Algorithms, Genetic Programming, etc.) and engineered algorithms (Expert systems, Cognitive systems, etc.). Hybrid types, such as algorithms engineered to evolve [28] and those, which evolve capability to do engineering are also possible. In biological domain, such mixed types can also be a result of crossbreeding between genetically engineered and naturally occurring organism.

For the purposes of our work it is important to establish clear definitions of what makes a sample artificial or natural as many samples will combine properties of both. Well engineered designs are capable of adaptation and some evolved systems are capable of engineering. For example, Shapiro argues that we observe natural genetic engineering in evolution: "… much of genome change in evolution results from a genetic engineering process utilizing the biochemical systems for mobilizing and reorganizing DNA structures present in living cells." [29]. We will define engineered samples as those which include any contributions from an intentional agent such as a human engineer, a definition which we hope makes it clear that it excludes natural evolution which is an intelligent [30] and powerful, but not purposeful or intentional [31], optimization process [32].

Finally, the possibility remains that a third type of an algorithm, one outputting random bits will also hit the target string [33], but as the size of the bit string grows such an algorithm would require an exponential amount of computational resources. Random algorithms could correspond to appearance of living forms by chance in some parts of the multiverse due to availability of necessary probabilistic resources [34]. It may happen if the Everett's many-worlds interpretation of quantum physics [35] is true or if a simple algorithm is used to generate every possible universe [36] leading to generation of every conceivable string in some universe, but as we are looking for a generic procedure to evaluate samples from particular universes, random algorithms could be safely ignored.

## 3. Distinguishing Naturally Evolved Life from Engineered Life

Analyzing properties of a particular evolutionary algorithm may allow us to discover features, which can be used to distinguish between engineered and evolved organisms. For one, we know that evolution takes a very long time to work so if we learned that only a limited amount of time was available for the formation of a complex sample that would indicate that it was not a product of natural evolution. Also, some features have not been found in natural systems and so their inclusion may indicate that engineering took place. For example, Minsky wrote: "Many computers maintain unused copies of their most critical "system" programs, and routinely check their integrity. However, no animals have evolved like schemes, presumably because such algorithms cannot develop through natural selection. The trouble is that error correction then would stop mutation--which would ultimately slow the rate of evolution of an animal's descendants so much that they would be unable to adapt to changes in their environments." [37].

Many respected scientists speak about apparent difficulty in distinguishing between natural and engineered systems. For example, Shapiro says: "It is very important to recognize that living cells resemble man-made systems for information processing and communication in their use of mechanisms for error detection and correction." [29]. Similarly, Dawkins, while speaking about evolved systems says: "Biology is the study of complicated things that give the appearance of having been designed for a purpose" [38] and continues "We may say that a living body or organ is well designed if it has attributes that an intelligent and knowledgeable engineer might have built into it in order to achieve some sensible purpose… any engineer can recognize an object that has been designed, even poorly designed, for a purpose, and he can usually work out what that purpose is just by looking at the structure of the object." [38]. More generally, Minsky addresses the need to change our thinking regarding teleological explanations: "We now can design systems based on new kinds of "unnatural selection" that can exploit explicit plans and goals, and can also exploit the inheritance of acquired characteristics.  It took a century for evolutionists to train themselves to avoid such ideas--biologists call them 'teleological' and Lamarckian'--but now we may have to change those rules!" [37]. Because evolution is a powerful optimization process it is capable of producing designs (springs [39], gears [40], compasses [41], Boolean logic networks [42], digital codes [43]) which are just as complex as those produced by intelligent agents, meaning that any test designed for detecting intelligence via examination of artifacts will fail to distinguish between the causal source [44].

The difficulty in assigning a sample to an origination process is exacerbated by the fact that most observed evidence is equally likely to support either synthetic or natural origins hypothesis. To see that let's compare observations of certain properties in naturally evolved biological organisms with similar observations but from engineered organisms or software: DNA similarities between organisms indicate that later samples evolved from earlier ones (ex. homo sapiens evolved from homo erectus), but code similarities between different releases of a software project indicate that the code was reused (ex. Windows NT and Windows XP). Poor design in nature can be explained by the fact that evolutionary process has no foresight (ex. blind spot), but poor design in engineered systems can be explained by incompetence of the engineer (ex. Toyota break problems). Vestigial organs in some animals (ex. wings of flightless birds) are well explained by deducing that the specie is in the process of adapting to a changed environment, but in the world of engineering outdated features are frequently observed because it may be costly to redesign the system to remove them (ex. ashtrays on airplanes) or to keep the system backwards compatible. Animals evolved ability to adapt to the changing environments (ex. seasonal fur change), but software is frequently designed to be adaptable to user preferences (ex. Netflix learning what movies you like).

Similar analysis can be applied to other evidence frequently used to justify attribution of samples to only a single hypothesis. It is important to note that this dual explanation for evidence is symmetric, it works both ways so "classical" evidence of engineering has a well-fitting explanation in naturalistic evolution and wise versa. Figure 1 illustrates why it may be difficult to distinguish natural and engineered specimens via simple observation.

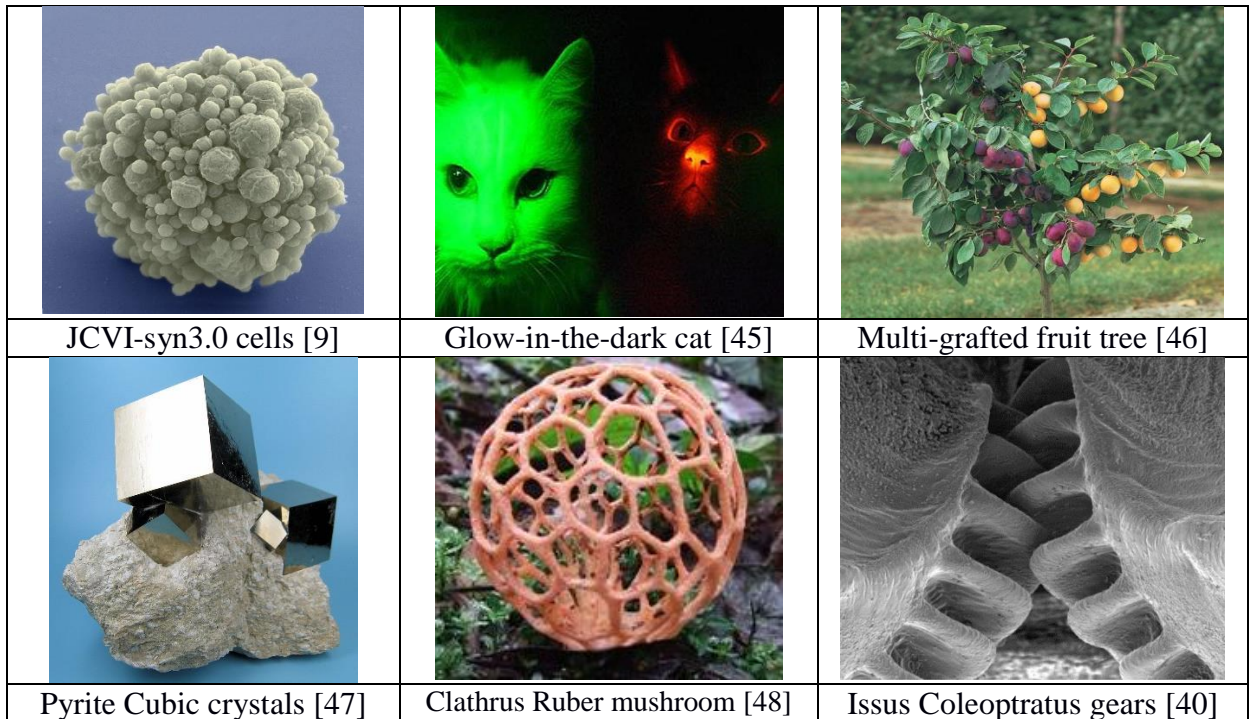

Figure 1: Engineered (top row) and natural samples may be difficult to separate out.

### 3.1 GMO Detection Methods

In order to comply with recent GMO regulations a number of techniques have been proposed to perform necessary analysis [17-19]. Many protein and nucleic acid-based detection methods have been developed and used for GMOs identification and quantification. Such techniques typically rely on direct matching of samples to available reference materials stored in databases of known GMOs, which might include sequence information of exogenous inserts as well as endogenous reference genes [49]. Such methods of direct matching do not work for undisclosed modifications.

### 3.2 Unevolvable Elements

An interesting direction in forensic investigation of origins of biological samples is study of Unevolvable Elements (UE). Such elements are components of the sample that could not arise via evolutionary process because all precursor elements do not improve or even lower fitness of the organism, preventing module (code fragment) in question from arising. We distinguish two types of such elements: those, which decode to a meaningful plaintext and are too long to happen by chance and those, which represent narrow targets in the space of possible solutions, surrounded by broad moats of negative fitness. While unevolvable elements of the first type are well documented [50, 51], existence in the real-world of elements of the second type remains an open question. Let us examine each type of UE, and review some examples of each.

In many cases, genetic engineers behind the project have no reason to hide their contribution and in fact may be interested in making sure that the organism is labeled in such a way that it is obviously seen as synthetic, for example with watermarks [52]. Labeling is also useful to make it possible to trace an organism's descendants to the originator, which is very important, for example in case of patent disputes [53]. Such labeling may take a form of a digital signature, or plain text metadata such as "Made in USA" text insertion.

Meaningful [33] text encoded in DNA on purpose or left there by mistake during design process (as comments or inactive code) could be detected and extracted [51]. In fact, over the last few decades scientists have inserted text messages into natural living organisms [54], GMOs [55] and synthetic life forms [52]. Such messages range in length from a few symbols such as "E=MC2" [54] to full text of books [56] with complete archival systems in the works [57]. The actual encoding and decoding process is beyond the scope of this paper, but the interested reader is advised to read a survey of the topic by Beck et al. [51]. As long as the length of the discovered message is not trivial an investigator can conclude that engineering took place and the organism is not 100% natural. Efforts to find such text [58, 59] preceded the ability of scientists to insert such messages. It seems that in general a search for signs of engineering in biological (genomic) information of any unattributed biological sample is just as reasonable as SETI search of astronomical data. In particular with any samples acquired from extraterrestrial sources such Biological SETI [60] should be a recommended first step.

The main challenge comes from recognizing text as "meaningful" particularly in cases of non-human engineers. Many attempts have been made to formalize "meaningful" to represent "the value of a message as the amount of mathematical or other work plausibly done by its originator" [61]. Measure variants proposed under such names as "potential" (Adleman [62]), "incomplete sequence" (Levin and V'jugin [63]), "hitting time" (Levin [64]), "sophistication" (Koppel [65]), and "intelligence based complexity" Yampolskiy [33] are best known as logical depth [61] of a string. Bennet describes this concept as follows: "Of course, the receiver of a message does not know exactly how it originated; it might even have been produced by coin tossing. However, the receiver of an obviously non-random message, such as the first million bits of pi, would reject this "null" hypothesis, on the grounds that it entails nearly a million bits worth of ad-hoc assumptions, and would favor an alternative hypothesis that the message originated from some mechanism for computing pi. The plausible work involved in creating a message, then, is the amount of work required to derive it from a hypothetical cause involving no unnecessary, ad-hoc assumptions. It is this notion of the message value that depth attempts to formalize." [61]. Similarly, Gurevich, describes a step-by-step process for what he calls "impugning randomness" [66] a method for distinguishing purposeful from accidental.

By analogy with the SETI approach a search for artificiality and cognitive universals could take place instead, with statistical abnormalities and non-randomness being used to detect language-like patterns [67, 68]. To avoid ambiguity it is desirable to find patterns which are 1) highly statistically significant and 2) in addition to exhibiting hallmarks of artificiality such as "symbol of zero, the privileged decimal syntax and semantical symmetries" are inconsistent in principal with any natural process be it Darwinian or Lamarckian evolution [60].

In adversarial scenarios, such as illegal utilization of GMOs, genetic engineers might be interested in hiding their contribution to the design of the organism, either by explicitly erasing all evidence or at least by making its detection difficult if not impossible without privileged information by relying on Stegonography [69] or deniable cryptography [70, 71]. Deniable cryptography produced by encoding and combination of multiple plain texts is not very efficient in terms of size of cipher text and would produce large segments of DNA with no discernable meaning, something akin to "junk DNA". However recent research suggest that such DNA segments are actually very meaningful and language like [72] and might contain historic record of modules which have evolved in previous environments and might be useful in the future if environmental conditions return to the previously seen state or for control gene expression. In case unknown engineers are suspected originators of the organism the text may be encoded using some unknown coding/language [73] so it might be a worthy idea to check Schelling Point [74, 75] passwords [76] such as digits of π, prime numbers, Fibonacci numbers, etc.

In organisms with no DNA code or if only external observation of the sample is possible we may be interested in investigating presence of unevolvable elements of second type - functional modules which could not arise via the process of mutation with natural selection due to low-fitness moats around such designs. A low-fitness moat does not just prevent evolution of the module from components; it also precludes its appearance as a reduction from a more complex module. Darwin himself put it as follows: "If it could be demonstrated that any complex organ existed, which could not possibly have been formed by numerous, successive, slight modifications, my theory would absolutely break down." [77]. Such modules can happen by random chance only if the number of involved parts is very small, so a component with a significant number of diverse parts is unlikely to arise by chance alone.

Existence of low-fitness moats, in complicated domains, such as biology is an open question we would like to see addressed. It is possible that they don't exist or are very rare. The argument is that the search space is so vastly high-dimensional (e.g. 3 billion base pairs in human DNA) that it is unlikely that there is literally no route through this 3-billion-dimensional space to any particular high-fitness point or region. There are similar arguments right now in deep learning about why stochastic gradient descent in large networks of millions of connections (i.e. dimensions) does not seem to be getting caught in local optima to the extent we might expect. It appears many of these "local optima" are actually saddle points and not optima after all, perhaps genome space is similar.

Consequently, we propose a challenge to the synthetic biology community to purposefully design and produce an organism with an unevolvable component, which meets Darwin's criteria for falsifying his theory. Can such a feat be accomplished? Can it be mathematically proven that a particular design is not evolvable, or at least statistically very unlikely? We believe those are important questions to be answered by genetic engineers and which would reconfirm falsifiability of the theory of evolution [78]. If unevolvable elements don't exist, every design can be naturally occurring and so it is not possible to distinguish between natural and synthetic origins, otherwise the presence of unevolvable elements can be used to proof that engineering took place.

A strong connection exists between unevolvable elements of the first and second type. A meaningful text may represent a blueprint for constructing an unevolvable organ and an unevolvable biological module can be reduced to a complex and meaningful informational pattern. By analogy with AI-Completeness [79, 80] we propose the concept of Intelligence-Completeness (I-Completeness) to indicate that certain elements are not evolvable and require intelligence to be constructed. I-Complete artifacts could be reduced to other representations (text, drawing, 3D model, organism, etc.) without losing its distinctive origination signature from the purposeful engineering process.

What distinguishes I-Completeness from AI-Completeness is that AI-Complete systems have no restrictions on how they can be constructed, while objects with the I-Complete property, from the definition, cannot be products of an evolutionary process. Consequently, our challenge of constructing an artificial unevolvable biological organ is equivalent to the problem of proving some problem I-Complete. From this first, hypothetical, case other problems would be shown to be I-Complete via a series of reductions, which is a method well-known in the theoretical computer science community [81].

AI-Completeness was first established [80, 82] as the property of passing the Turing Test (TT) [83], with other problems shown to be AI-Complete via reductions from the TT. Perhaps we can rely on the same problem for proving I-Completeness, since engineering of synthetic life requires at least human level intelligence, and that is exactly what is being detected by TT. One possibility is to take verbatim text from someone passing TT and to encode it in organism's DNA, with questions from the test corresponding to specifications and answers to meaningful information. Existence of area specific TTs in domains such as art and poetry [84] suggest that we can also produce a restricted TT for the domain of genetic engineering and encode any unevolvable element descriptions as answers to questions asking to describe such structure.

**3.3 Forensic Evidence from the Code**
In general, as long as statistical properties of samples produced by a particular algorithm can be captured, another algorithm can simulate them on purpose, essentially spoofing behavior of the original algorithm [85]. If fact the statistical model describing the samples is an engineered algorithm for generating equivalent sample distribution be it from an evolutionary process or any other type of algorithm. Engineered algorithms are capable of both simulating natural evolution and using it as a module in achieving their goals [86]. In principal, an engineered algorithm can produce any computable distribution and so can an evolutionary algorithm with infinite computational resources, making both types of algorithms universal and claims of particular origin of samples unfalsifiable due to unlimited power of either approach. Consequently, we can never have 100 percent certainty as to the origination algorithm, only probabilistic estimates. Of course our analysis applies only to post-factum observations of collected samples. If we have a chance to observe and analyze sample generator at work we can be certain as to the used process.

**4. Designometry - Generalization of the Proposed Analysis**
Forensic investigators studying an explosive device, professor looking at plagiarized programming project, art experts examining a potential forgery, and numerous other professionals find themselves in a situation where they need to infer information about the engineer/designer/author of a product/object/text in the absence of direct access to the agent only in possession of the agent's output. For example, depending on the domain the process of making such inference may be called forensic analysis [51], stylometry [87], historiometrics [88] or behavioral profiling [89, 90]. Regardless of the subdomain of inquiry, the goal of the generalized process we will call *Designometry,* is to discover "signature" of the originator in the artifact and from it to identify the agent responsible or to at least learn some properties likely to be necessary for the design process, which had to take place to produce the artifact. Designometry could be widely applied to both non-biological and living artifacts, which are products of intentional construction. The field commonsensically includes such subdomains as:

- **Artimetrics** – identifies software and robots based on their outputs or behavior [91, 92].
- **Behavioral Biometrics** – Quantifies behavioral traits exhibited by users and uses resulting feature profiles to verify identity [93]. Examples of analyzed artifacts may include text, art, as well as direct or indirect human-computer-interaction [94].
- **CAPTCHA (Completely Automated Public Turing test to tell Computers and Humans Apart)** – obtains input from an agent and classifies producing agent as human or artificial [95, 96].

And could itself be seen as a sub-branch of Intellectology a field proposed to "study and classify design space of intelligent agents, work on establishing limits to intelligence (minimum sufficient for general intelligence and maximum subject to physical limits), contribute to consistent measurement of intelligence across intelligent agents, look at recursive self-improving systems, design new intelligences (making AI a sub-field of intellectology) and evaluate capacity for understanding higher level intelligences by lower level ones" [97, 98].

Next, we will give a few current illustrations, which would fall under the heading of designometry. For example, stylometry of text relies on statistical analysis of "vocabulary richness, length of sentence, use of function words, layout of paragraphs, and key words" [99] to determine gender, age [100], native language [101], personality type [102] and even intelligence of a human author or comparable properties of or artificially intelligent text generator [103]. In general, it seems it is possible to estimate scientific knowledge and minimum intelligence necessary to produce, or at least duplicate, a particular artifact by analyzing its complexity, prerequisite components and evidence of tools used in the production, be it a material object or an abstract algorithm/data [33, 65]. This doesn't imply that anyone with the required level of intelligence would be able to produce artifact under consideration, just that someone below that level would fail to do so. Reader is encouraged to read about fascinating designometric analysis of Antikythera mechanism [104], Egyptian pyramids [105] or Stuxnet Virus [106] for some famous examples of such efforts.

As for anticipated future applications of designometry, one example could be given from the domain of AI Safety, Yampolskiy writes about an artificial superintelligent system confined to a restricted environment [107], which attempts to learn about the nature of its designers/programmers by inspecting its own source code: "… the AI will have access to covert sources of information such as its own hardware and software and could analyze its design and source code to infer information about the designers. For example analysis of the source code may reveal to the AI that human programmers are slow (based on the file modification dates), inefficient (based on code redundancy), illogical (based on bugs in the code), have bad memory (based on the long and descriptive variable names), and don't think in code (based on unnecessary comments in the code)" [108]. Another interesting application of designometry would be to the problem of determining if the environment in which an agent (human or artificial intelligence) finds itself is natural or engineered. This has important applications in the domains of AI Safety [109], self-locating beliefs [110], life choices [111] and general philosophy [23]. Such capacity would be particularly timely as our ability to create realistic virtual worlds is improving exponentially [112]. Finally, we foresee great utilization in domain of steganography [113] detection and general forensic analysis.

Open problems in designometry include consolidation of analysis methods from sub-field specific domains, as well as development of generalized tools and tests to be used without modification in novel domains of investigation. Man-made [114, 115], alien-made [116] and artificial object detection, and exhaustive understanding of types of information which could be inferred about the originator from the artifact are all current examples of research directions in designometry. It may also be useful to be able to tell if two designs were engineered by the same agent or if an agent reused parts from another design. It is also highly likely that this process could be automated via machine learning as has been demonstrated by recent work in software designometry [117].

## 5. Most Life in the Universe has Engineered Origins

Inspired by Bostrom's statistical argument for our universe being an engineered one [23] we suggest a similar argument but in realm of biology. Estimating the distribution, in the universe, of synthetic life versus naturally occurring life it is likely that designed life (biological robots of any complexity produced by early alien civilizations) is the significantly more common default case. Others have made similar observations, for example Dick: "…cultural evolution may have resulted in a postbiological universe in which machines are the predominant intelligence""… this means that we are in the minority; the universe over the billions of years that intelligence has had to develop will not be a biological universe, but a postbiological universe" [118], or Schneider specifically on high intelligence agents: "… it may be that [Biologically Inspired Superintelligent Aliens] are the most common form of alien superintelligence out there." [119]. Similarly, Makukov et al, state: "… at the current age of the Galaxy it might be even more probable for an intelligent being to find itself on a planet where life resulted from directed panspermia rather than on a planet where local abiogenesis took place, and the Earth is not an exception from that. This is not to say that the view that terrestrial life originated locally is flawed. But subscribing largely to this view and dismissing the possibility that terrestrial life might not be a first independent generation in the Galaxy is probably nothing but a manifestation of geo-anthropo-centrism (inappropriately armed with Occam's razor)." [120].

Unless evidence to the contrary exists, a given life form is statistically more likely to have its origins as a product of engineering and so our priors should be adjusted accordingly. This type of reasoning also applies to Earth, we are also likely to have our origins as synthetic life as suggested by the theory of directed panspermia [22], or seeding [121] or some other similar variants [122, 123]. In fact approximate probability of being produced by unaided laws of physics rather than engineering is equal to 1 divided by the total number of self-reproducing biological robot species all the generations of intelligent beings around the universe have ever produced. A result which in our estimate (based on Drake's equation [124]) tends to approach zero as the age of the universe increases. In general, as the universe ages, the chance of any life form being an original evolved form rather than second or later generation design approaches zero. It is important to note that our statistical argument applies only to origins of life, not to the process of speciation, which of course is well explained by the theory of evolution/adaptation. Additionally, theory of evolution doesn't make any claims regarding origins of life.

Assuming that in our future we will seed thousands if not millions of such robot colonies (which in turn may do the same), we can observe that the common problem of origins attribution would show up on many planets (this would also happen under the many-worlds interpretation of quantum mechanics and as a result of robots developed by space aliens). We may refer to it as the Many Darwins Problem (a "Darwin" per seeded planet).

Further, let us consider a thought experiment; we shall call the Robot Planet Problem[1]. Suppose at some point in our future we design a very advanced humanoid (biological) self-replicating robot with the goal of exploring distant planets. We send a group of such robots on a long-term mission to a star known to be orbited by a number of earth-like planets [125]. Our goal may be to establish a permanent base on one of more of such planets, to reserve its resources for us, in case competing alien species may have interest in the same solar system. We would also like to make said planets habitable for human beings and to instruct our robots to await contact from its human masters. The robots are, of course, designed to be adaptable to variations in their future environment and have a general level of intelligence comparable to that of humans.

It may be possible to make them superintelligent [98, 108, 119, 126], but it is probably not a rational thing to do as such robot may present danger to us and others and are harder to control [127]. Also, providing robots with very specific goals may produce undesirable side-effects and may not work well in a large number of planets with unknown conditions. Perhaps our instructions to them will be something like: "Reproduce to a number sufficient to obtain full control of your host planet, make it habitable for yourself and your masters and await arrival of your designers". A number of less important instruction can be provided, such as: maintain good condition of each robot, establish a rule of law, do not destroy other robots, etc. It is possible that the planets in question may already contain some forms of life, but probably not highly intelligent life, so additional instructions may be provided to preserve local biodiversity.

As a significant amount of time passes on the Robot Planet, group's mission is probably going to progress fairly well with construction of necessary infrastructure, increase in population and development of sophisticated local culture and religious tradition centered around its human masters. At some point, most or all robots would have no direct knowledge of their human masters. Advanced science would also likely be conceived by that time. At this point it is extremely likely that a "Robot Darwin" would appear, who would criticize idea of human masters as an irrational belief and propose a naturalistic explanation for the inhabitants of the robot planet not too different from the theory of evolution. Since the robots were designed with ability to adapt to their new environment sufficient evidence for evolution would be found and it would quickly become a dominant and very reasonable explanation for the origins of the robot colony, in the light of ideas presented in this paper.

## 6. Conclusions
In this paper, we have suggested a design for an experiment in which engineered life is as likely as natural one by normalizing priors. The experiment is intended to test the current assumption that it is possible to determine if a given sample is produced by natural evolution while also allowing us to investigate detectability of genetically modified and fully synthetic life forms, which are quickly becoming common due to the latest advances in genetic engineering. With some thought experiments we have shown that most current life is statistically more likely to have synthetic origins and shown how such theory could be tested by translating the problem to the domain of computer science - algorithms and data. Importantly all investigated theories have fully naturalistic explanations and are completely scientific.

---
[1]We are well aware of the Futurama episode "A Clockwork Origin" (Episode 6, Season 9) with a similar plot.

In a theoretical case of unlimited resources (mostly time [128], but also multiverses) it is not possible for some samples to tell which type of algorithm is responsible for producing collected samples as all investigated algorithms are universal in a sense that they can eventually produce any pattern. The suggested analysis is also broadly applicable to biological and non-biological samples, essentially everything we can represent as a binary string.

Developments in synthetic biology and evolutionary robotics raise a number of ethical, biosafety and security issues. In addition to potential development of novel deadly pathogens [129], genetically modified humans [130] and other organisms, we are also facing a potential runaway evolutionary process. An outcome of such process could be appearance of dangerous and potentially superintelligent robots [131], which may cause human extinction in the same way that a large number of previously existing species went extinct because of appearance of an intellectually superior specie - Homo Sapiens.

We have reviewed a number of cases in which it is possible, as a result of forensic analysis, to conclusively state that a collected sample has been engineered rather than occurred naturally. Such telltale signs may include: complexity in the absence of probabilistic resources, watermarking, programmers comments, multilevel encoding [132], support for future features, evidence of degradation from the original design, engineer's signature, etc. It may even be possible for intelligent agents to perform this analysis on themselves to discover their origins. Synthetic life forms which may be discovered in the wild will be interesting to study because they can have a number of features not found in naturally occurring ones, such as: backdoor control mechanisms, hidden capabilities, previously unseen features, etc. Studying designed systems may also leak information about the engineer(s) behind the design. Methods to do so are of interest to forensic investigators, SETI scientists, stylometry practitioners, and exobiologists. Finally, it is very important to note that confirmed detection of synthetic life, even in the wild, doesn't prove any non-naturalistic notions be it god(s), creationist myths, or religion, only that engineering took place.

**Acknowledgements**
Author is thankful to Yana Feygin, Alexey Melkikh, Susan Schneider, Gennadiy Mirochnik and Kenneth Stanley for valuable feedback on this paper.